\documentclass[9pt,twocolumn,twoside]{opticarx}
\setboolean{shortarticle}{false}
\usepackage{braket}
\title{VECSEL systems for generation and manipulation of trapped magnesium ions}

\author[1*]{S. C. Burd}
\author[1]{D. T .C. Allcock}
\author[2]{T. Leinonen}
\author[2]{J. P. Penttinen}
\author[1]{\\D. H. Slichter}
\author[1]{R. Srinivas}
\author[1]{A. C. Wilson}
\author[1]{R. J{\"o}rdens}
\author[2]{M. Guina}
\author[1]{\\D. Leibfried}
\author[1]{D. J. Wineland}

\affil[1]{Time and Frequency Division, National Institute of Standards and Technology, 325 Broadway, Boulder, Colorado, 80305, USA}
\affil[2]{Optoelectronics Research Centre, Tampere University of Technology,
PO Box 692, FIN-33101 Tampere, Finland}

\affil[*]{Corresponding author: shaun.burd@nist.gov}

\dates{Compiled \today}

\ociscodes{(140.5960) Semiconductor lasers; (140.3515) Lasers, frequency doubled; (020.3320) Laser cooling; (270.5585)  Quantum information and processing; (300.6320) Spectroscopy, high resolution;  (300.6350) Spectroscopy, ionization; (300.6520) Spectroscopy, trapped ion;}

\doi{\url{http://dx.doi.org/10.1364/ao.XX.XXXXXX}}

\begin{abstract}
  Experiments in atomic, molecular, and optical (AMO) physics rely on lasers at many different wavelengths and with varying requirements on spectral linewidth, power, and intensity stability. Vertical external-cavity surface-emitting lasers (VECSELs), when combined with nonlinear frequency conversion, can potentially replace many of the laser systems currently in use. Here we present and characterize VECSEL systems that can perform all laser-based tasks for quantum information processing experiments with trapped magnesium ions. For photoionization of neutral magnesium, 570.6$\,$nm light is generated with an intracavity frequency-doubled VECSEL containing a lithium triborate (LBO) crystal for second harmonic generation. External frequency doubling produces 285.3$\,$nm light for resonant interaction with the $^{1}S_{0}\leftrightarrow$ $^{1}P_{1}$ transition of neutral Mg. Using an externally frequency-quadrupled VECSEL, we implement Doppler cooling of $^{25}$Mg$^{+}$ on the 279.6$\,$nm $^{2}S_{1/2}\leftrightarrow$ $^{2}P_{3/2}$ cycling transition, repumping on the 280.4$\,$nm $^{2}S_{1/2}\leftrightarrow$ $^{2}P_{1/2}$ transition, coherent state manipulation, and resolved sideband cooling close to the motional ground state. Our systems serve as prototypes for applications in AMO requiring single-frequency, power-scalable laser sources at multiple wavelengths.
\end{abstract}

\setboolean{displaycopyright}{false}

\begin{document}

\maketitle
\thispagestyle{fancy}

\ifthenelse{\boolean{shortarticle}}{\ifthenelse{\boolean{singlecolumn}}{\abscontentformatted}{\abscontent}}{}

\section{Introduction}

\begin{figure}
\centering
\includegraphics[width=0.9\linewidth]{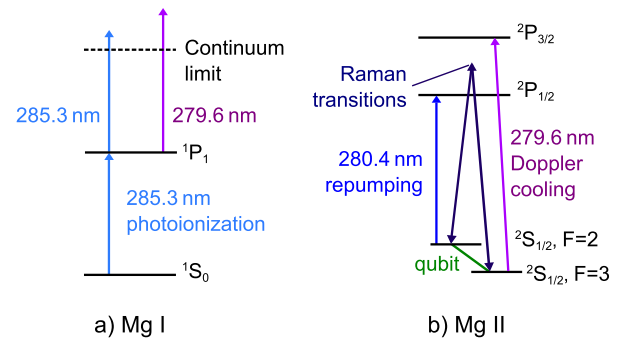}
\caption{Energy levels of a) neutral magnesium relevant for photoionization and b) $^{25}$Mg$^{+}$ relevant for Doppler cooling, repumping and stimulated-Raman transitions.}
\label{fig:Energylevels}
\end{figure}

Trapped ions provide a versatile experimental platform for research in quantum information processing \cite{Didi2003,Steane1997,James1998,Blatt2008}, quantum simulation \cite{Britton2012,Blatt2012,Monroe2015,Schneider2012}, and precision metrology \cite{Rosenband2008, Kotler2014}. Singly-ionized magnesium has several desirable properties for ion trapping experiments. It has a cycling transition, which implies that only a single laser is required for Doppler cooling. High motional frequencies can be achieved with relatively low trap drive voltage due to the low atomic mass. The isotope $^{25}\,$Mg$^{+}$, with nuclear spin I$\,$=$\,$5/2, has a series of intermediate magnetic field, hyperfine "clock" qubits which are, to first order, insensitive to magnetic field fluctuations. The $\sim\,$$1.7\,$GHz hyperfine splitting of these qubits facilitates manipulation with readily available, commercial microwave electronics \cite{Ospelkaus2011}. Magnesium ions have been used in experiments such as the one of the earliest demonstrations of laser cooling \cite{Wineland1978} and microwave-driven two-qubit quantum logic gates \cite{Ospelkaus2011}. However, generating the necessary high-power UV light presents a number of technical challenges. 

 For Mg$^{+}$ ion trapping applications, resonant two-photon ionization is generally used instead of direct excitation to the continuum at $7.646\,$eV ($162.6\,$nm) \cite{Masden2000}. As shown in Figure \ref{fig:Energylevels}\textcolor{blue}{a}, the photoionization (PI) laser, on resonance with the 3s$\,^{1}S_{0}\leftrightarrow \,$ 3p$\,^{1}P_{1}$ transition of neutral magnesium at $285.2\,$nm, excites the atom. Photons from the PI laser or from another laser near $280\,$nm (for Doppler cooling Mg$^{+}$ ions) then have sufficient energy to ionize the atom. The energy levels employed for Doppler cooling, repumping and quantum state manipulation are shown in Figure \ref{fig:Energylevels}\textcolor{blue}{b}. The 3s$\,^{2}S_{1/2}\leftrightarrow \,$ 3p$\,^{2}P_{3/2}$ transition provides a cycling transition for Doppler cooling and state detection. For quantum information experiments with $^{25}$Mg$^{+}$, pairs of appropriate Zeeman sublevels in the 3s$\,^{2}S_{1/2}\ket{F=3}$ and 3p$\,^{2}S_{1/2}\ket{F=2}$ manifolds can serve as qubits. Coherent qubit manipulation can be achieved with stimulated-Raman transitions, where the frequency difference between two Raman laser beams is set close to the qubit frequency (Figure \ref{fig:Energylevels}\textcolor{blue}{b}). Stimulated-Raman transitions allow coupling of internal degrees of freedom of the ions with their motional degrees of freedom \cite{Didi2003} which is essential for implementing quantum logic gates and sub-Doppler cooling. Resolved sideband cooling can be used to cool trapped ions close to the motional ground state \cite{Didi2003}. Preparation of a quantum system in a specific motional state with high probability is required for many quantum information experiments and for trapped-ion atomic clocks \cite{Rosenband2008}. During resolved sideband cooling, it is necessary to repump the ion back to the initial hyperfine state. This can be accomplished efficiently using a laser resonant with the  3p$\,^{2}S_{1/2}\leftrightarrow \,$ 3p$\,^{2}P_{1/2}$ transition.      

Magnesium ion trapping experiments have been performed traditionally with frequency-doubled, traveling-wave dye lasers. More recently, frequency-quadrupled ytterbium fiber lasers \cite{Hemmerling2011,Friedenauer2006} have been used to generate the light for Doppler cooling and qubit manipulation. However, these fiber lasers operate near the long wavelength edge of the amplification curve; thus they are typically limited to about 2$\,$W in the infrared (IR) at 1118$\,$nm. To the best of our knowledge, commercial fiber laser systems are not currently available that can reach either 1120$\,$nm for repumping on the $^{2}S_{1/2}\,\leftrightarrow\,^{2}P_{1/2}$ transition or 1141$\,$nm for photoionization. Frequency-quadrupled diode laser systems are available that cover all required wavelengths for magnesium ion experiments \cite{Zhang2013}. These systems are currently limited to about 2$\,$W of IR power if a tapered amplifier is used to amplify the diode seed laser. To achieve higher power, amplification of a single-frequency seed laser can be accomplished with a Raman fiber amplifier \cite{Feng2009}. High-power, widely tunable optical parametric oscillator (OPO) systems are also available, but require a single-frequency pump source.

Here we demonstrate a laser system for magnesium ion trapping experiments based on vertical external-cavity surface-emitting laser (VECSEL) technology. VECSELs exhibit many desirable properties for applications in high-precision spectroscopy \cite{Burd2015}. The gain material can be designed for emission over a broad range of wavelengths with demonstrations at 390$\,$nm in the ultraviolet, and at multiple wavelengths from 674$\,$nm in the visible to 5$\,$µm in the infrared \cite{Kuznetsov2010}. The VECSEL geometry allows single-frequency emission with close to diffraction-limited output beam quality and a tuning range of tens of nanometers. External or intracavity harmonic generation extends the attainable wavelength range further into the ultraviolet. The VECSEL architecture is intrinsically amenable to power scaling. Single-frequency VECSELs with output powers of up to 23.6$\,$W in the near IR have been reported \cite{Zhang2014}. Recently, a 0.56$\,$W, 229$\,$nm source, from the fourth harmonic of a 916$\,$nm VECSEL, has been developed \cite{Kaneda2016}. VECSEL systems with external harmonic generation have been used to perform Doppler-free spectroscopy of neutral mercury \cite{Paul2011} and neutral rubidium \cite{Cocquelin2009}. Our systems consist of two laser designs. The first is an externally frequency-quadrupled VECSEL capable of generating 3.0$\,\pm\,$0.1$\,$W output power at 1117$\,$nm which can be wavelength tuned for either laser cooling, repumping or stimulated-Raman transitions of Mg$^{+}$ ions. The second system, used for photoionization, is an intracavity frequency-doubled VECSEL emitting at up to 2.4$\,\pm\,$0.1$\,$W (total power in two output beams) at 571$\,$nm and externally doubled to 285$\,$nm. Both systems use commercial multi-mode diode lasers as pump sources. 

\section{Laser setup and characterization}

\begin{figure}
\centering
\includegraphics[width=0.8\linewidth]{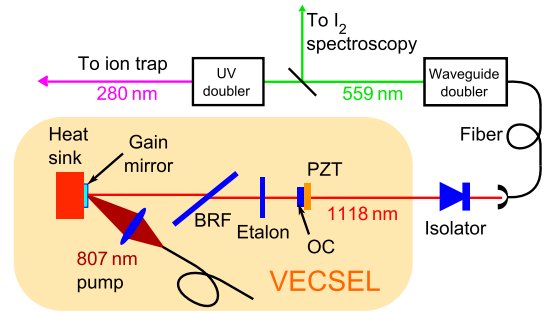}
\caption{Externally quadrupled, I-cavity VECSEL system (IC system) used for Doppler cooling, repumping and performing stimulated-Raman transitions. PZT Piezo transducer, OC Output coupler, BRF Birefringent filter}
\label{fig:Laser_Setups}
\end{figure}

\begin{figure}
\centering
\includegraphics[width=0.8\linewidth]{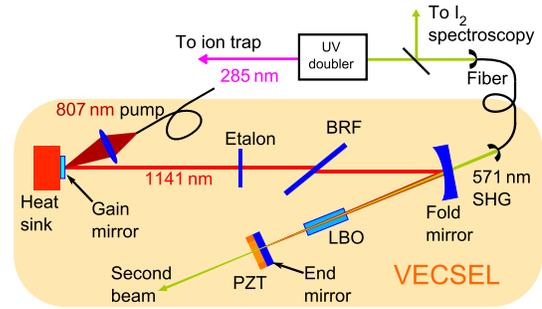}
\caption{V-cavity, intracavity frequency-doubled system (VC system) for photoionization. LBO lithium triborate, SHG second harmonic generation}
\label{fig:PI_Laser_Setup}
\end{figure}

The externally frequency-quadrupled, I-cavity VECSEL (IC system) is shown in Figure \ref{fig:Laser_Setups}. The bottom-emitting semiconductor gain chip, consisting of GaInAs/GaAs quantum wells strain compensated by GaAsP layers \cite{Ranta2011}, is mounted on a diamond heat spreader. A thermoelectric cooler (TEC) module is clamped between the gain chip mount and a water-cooled micro-channel cooler for thermal management and temperature control. The 12.5$\,$cm-long resonator has an I geometry consisting of the gain mirror and a 200$\,$mm  radius of curvature (ROC), 1.5$\,$\% output coupler. The end mirror is glued to a ring piezo-electric transducer (PZT) for controlling the cavity length. The gain chip is pumped with an 807$\,$nm multimode laser diode. The $1/e^{2}$ intensity pump spot diameter on the gain chip is roughly 400$\,$µm. A 3$\,$mm-thick quartz birefringent filter (BRF) plate, inserted at Brewster's angle, and a 1$\,$mm-thick yttrium aluminium garnet (YAG) etalon are placed in the cavity to ensure single mode operation and to provide wavelength selection. The etalon is positioned in a temperature controlled oven for frequency stability and wavelength tuning. The VECSEL is contained in an insulated enclosure for acoustic and thermal isolation.
With this system design we have achieved 3.1$\,\pm\,$0.1$\,$W of IR power at 1117$\,$nm with 21$\,\pm\,1$W of 807$\,$nm pump power (slope efficiency of 19$\,\pm\,1\,$\%). Light from the laser propagates through an optical isolator to prevent back reflection and is coupled into a commercial fiber-coupled quasi-phase matched lithium niobate waveguide doubler. We operate the doubler at second harmonic (SH) powers below 300$\,$mW to avoid potential damage to the waveguide. We obtain overall conversion of 1118$\,$nm to $559\,$nm of 22$\,\pm\,2$\% (including fiber coupling losses). SH light is coupled into a build-up cavity containing a $\beta$-barium borate (BBO) crystal for frequency doubling to 280$\,$nm. Details of the cavity design are given by Wilson \emph{et al.} \cite{Wilson2011}. With 1160$\,\pm\,40\,$mW in the IR, resulting in 260$\,\pm\,10\,$mW SH power, we generate 23$\,\pm\, 2\,$mW in the UV with a visible to UV conversion efficiency of 9$\,\pm\,1\,$\%. 

\begin{figure}
\centering
\includegraphics[width=1\linewidth]{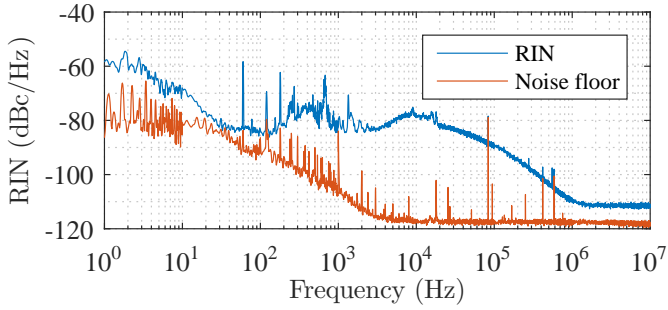}
\caption{Relative intensity noise spectrum at 280$\,$nm. The noise floor is the spectrum obtained with the laser blocked, all other experimental parameters the same.}
\label{fig:RB_RIN}
\end{figure}

The laser linewidth in the UV must be much narrower than the width of the relevant atomic transitions (typically tens of MHz) to be of use for precision spectroscopy or Doppler cooling. In addition, resonant doublers for generating UV light will convert laser frequency noise into intensity noise, which will degrade the fidelity of quantum manipulations performed using stimulated-Raman transitions \cite{Wineland1998}. Figure \ref{fig:RB_RIN} shows the relative intensity noise (RIN) spectrum of the IC system when producing 13$\,\pm\, 1\,$mW at 280$\,$nm. The experiment was performed using a photodiode connected to a fast Fourier transform analyzer, following the basic procedure described by Obarski and Splett \cite{Obarski2001}. The RIN in Figure \ref{fig:RB_RIN} noise is below -75$\,$dBc/$\mathrm{Hz}$ above 10$\,$kHz. 

\begin{figure}
\centering
\includegraphics[width=1\linewidth]{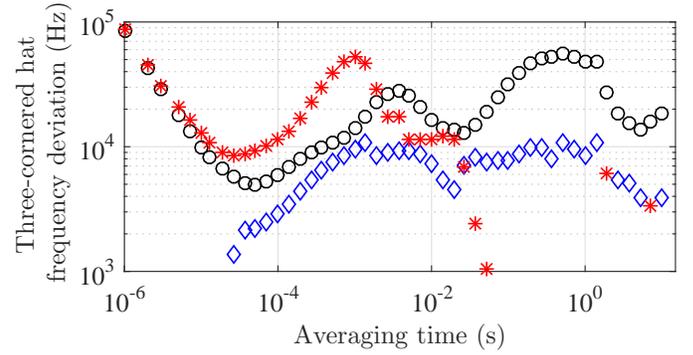}
\caption{Three-cornered hat frequency deviation (square root of the frequency variance) comparing the IC VECSEL system (blue diamonds) with a fiber laser (red stars) and an external-cavity diode laser (black circles).}
\label{fig:fig}
\end{figure}

\begin{figure}[t]
\centering
\includegraphics[width=1\linewidth]{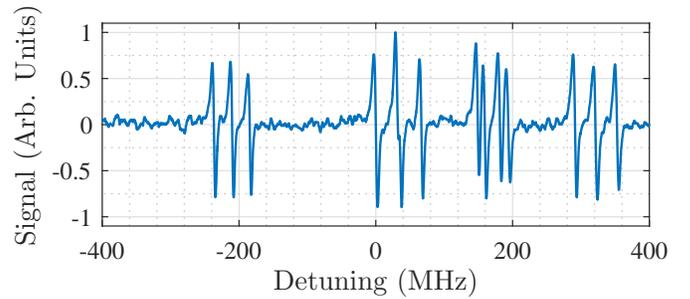}
\caption{Doppler-free scan of iodine lines about 560.4050$\,$THz using the intracavity-doubled VC system.}
\label{fig:ISpectrum}
\end{figure}

The "three-cornered hat" technique \cite{Gray1974} was used to measure the short-term frequency fluctuations of the IC VECSEL in the infrared. Beams from the VECSEL, a single-mode fiber laser, and an external-cavity diode laser were combined and directed onto a single photodiode. The lasers' frequencies were set sufficiently close together that time-domain measurements of the beat-notes of each pair of lasers could be recorded. The frequency of each laser was stabilized on slow time scales by locking to Doppler-free iodine spectroscopic features (closed-loop bandwith <$\,$2$\,$Hz), using a variation of the spectroscopic method described by Snyder \emph{et al.} \cite{Snyder1980}. The data were processed to give the frequency deviation of each laser at different averaging times as shown in Figure \ref{fig:fig}. The VECSEL exhibits a frequency deviation below $20\,$kHz over all averaging times. For laser cooling and repumping experiments, it was necessary to stabilize the laser frequency to reduce slow drifts (over minute time scales). This was achieved either by locking the VECSEL to Doppler-free spectroscopic features of molecular iodine or performing a frequency offset lock \cite{Castrillo2010} to an iodine-locked fiber laser. For stimulated-Raman transitions, the free running VECSEL frequency was sufficiently stable and active frequency control was not required.

The intracavity frequency-doubled VECSEL system for photoionization (VC system) is shown in Figure \ref{fig:PI_Laser_Setup}. The same gain chip mechanical mount and thermal management are utilized as in the IC system. The gain mirror consists of ten GaInAs quantum wells strain-compensated by GaAsP. The cavity has a V-geometry formed by the gain mirror, a ROC 100$\,$mm fold mirror and a plane end mirror. The mirrors are coated for high reflectivity from 1116$\,$nm to 1142$\,$nm and for high transmission at second harmonic wavelengths. A temperature stabilized 3$\times$3$\times$15$\,$mm lithium triborate (LBO) crystal, cut for type I phase matching, is used for intracavity second harmonic generation. A 1$\,$mm-thick YAG etalon and 3$\,$mm-thick quartz BRF inserted at Brewster's angle are used for wavelength selection. A PZT is used for fine adjustment of the cavity length. Two separate 571$\,$nm beams with roughly equal power exit the cavity through the fold and end mirrors. The laser can generate up to 1.2$\,\pm\,0.1\,$W in each output beam at 571$\,$nm when the gain mirror is cooled to 8$\,$$^{\circ}$C. The 571$\,$nm light is frequency-doubled to 285$\,$nm using a build-up cavity with a BBO crystal. For reduction of long-term frequency drifts, the VC laser can be locked to a Doppler-free molecular iodine transition. A scan over the Doppler-free iodine lines near 525.5040$\,$THz is shown Figure \ref{fig:ISpectrum}. From the locked error signal, we estimate a root mean squared (RMS) frequency deviation of 180$\,$kHz over 6 minutes with a lock-in time constant of 10$\,$ms. The short term frequency fluctuations were measured by tuning the laser onto resonance with a reference cavity (finesse of 250$\,\pm\,10$). From the electronic H{\"a}nsch-Couillaud error signal \cite{Hansch1980} we infer a laser linewidth of 50$\,\pm\,10\,$kHz using a detector bandwidth of 470$\,$kHz.

\section{Ion trapping and spectroscopy}

In our Mg ion trapping apparatus, a thermal beam of isotopically enriched $^{25}$Mg atoms is produced using an electrically heated oven. For photoionization, roughly 1$\,$mW of laser light at 285$\,$nm, from the VC system, is coupled into a solarization-resistant single-mode fiber \cite{Colombe2014} and focused to a calculated $1/e^{2}$ intensity diameter of 26$\,$µm through the trapping region of a surface-electrode RF Paul trap similar to that used by Ospelkaus et. al. \cite{Ospelkaus2011}. Light at 280$\,$nm for Doppler cooling is also applied to the trapping region. After allowing the oven to reach its steady-state operating temperature, we reliably and repeatably load single Mg$^{+}$ ions into the trap within 10$\,$s. 

\begin{figure}
\centering
\includegraphics[width=1\linewidth]{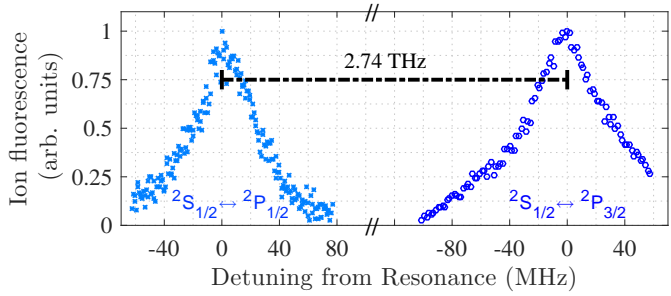}
\caption{Spectroscopy of the $^{2}S_{1/2}\leftrightarrow$$^{2}P_{1/2}$ and $^{2}S_{1/2}\leftrightarrow$$^{2}$P$_{3/2}$ transitions of a single trapped $^{25}$Mg$^{+}$ ion. The zero detuning points of the transitions from the $^{2}S_{1/2}$ to the $^{2}P_{1/2}$ and $^{2}P_{3/2}$ levels are relative to 1069.34$\,$THz and 1072.08$\,$THz respectively \cite{Clos2014}. The fine structure separation is approximately 2.74$\,$THz. Each spectrum is independently normalized.}
\label{fig:Spectroscopy}
\end{figure}

For spectroscopy of the $^{2}S_{1/2}\leftrightarrow $$^{2}P_{3/2}$ transition, the IC system was frequency-offset-locked to a fiber laser stabilized to a Doppler-free iodine line. UV light from the IC system is fiber coupled and focused onto the ion with a $1/e^{2}$ intensity diameter of approximately 24$\,$µm. The frequency of the light is shifted by adjusting the drive frequency applied to a double-pass acousto-optical modulator (AOM). At each frequency point, photons scattered by the trapped ion were collected with a high numerical aperture objective and counted with a photomultiplier tube (PMT). The experiment was repeated 200 times at each frequency point. The UV light from the VECSEL was pulsed on for $25\,$µs during each experiment so as not to significantly change the motional state of the ion when the frequency is blue detuned. Figure \ref{fig:Spectroscopy} shows a scan of the AOM frequency over the resonance. The linewidth was measured to be 69$\,\pm\,3\,$MHz. The residual width, above the natural linewidth of 41.8$\,$MHz \cite{Clos2014}, is likely due to residual Doppler broadening. To demonstrate laser cooling, the laser was detuned by $\sim$$10\,$MHz to the red of the line center. The ion could be confined in the trap with no change in ion lifetime compared to Doppler cooling with a fiber laser system. Decoherence-assisted spectroscopy \cite{Clos2014} was used to perform spectroscopy of the $^{2}S_{1/2}\leftrightarrow^{2}P_{1/2}$ transition. Figure \ref{fig:Spectroscopy} shows the resulting spectrum. The IC system can be tuned over the $\simeq\,2.74\,$THz fine structure splitting of $^{25}$Mg$^{+}$. 

\begin{figure}[t]
	\vspace{1.5mm}
\centering
\includegraphics[width=1\linewidth]{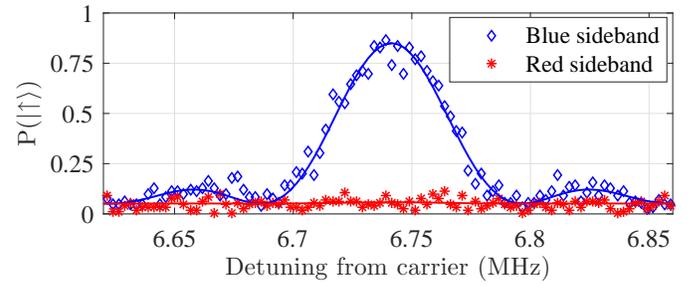}
\caption{The $\ket{\uparrow}$ state population of the first radial motional mode red and blue sidebands after ground state cooling resulting in $\braket{n}=0.008\,^{+\,0.013}_{-\,0.008}$.}
\label{fig:Side_band_cooling}
\end{figure}

\begin{figure}[t]
\centering
 \includegraphics[width=1\linewidth]{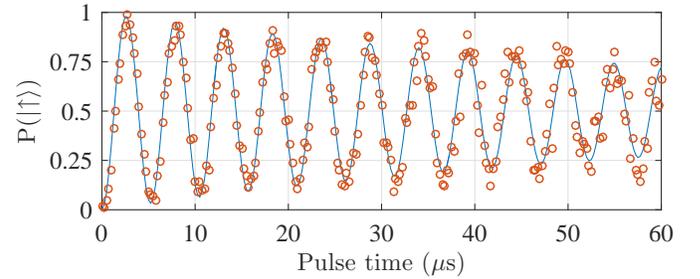}
\caption{Rabi flopping on the $\ket{\downarrow}\leftrightarrow\ket{\uparrow}$ carrier transition for a ground state cooled $^{25}$Mg$^{+}$ ion. The solid line is a exponentially decaying sinusoid, fitted to the data to give a Rabi frequency of 191.1$\,\pm\,$ 0.1$\,$kHz and a $1/e$ decay time of 77$\,\pm\,$3$\,$µs.}
\label{fig:Flopping}
\end{figure}

For experiments involving stimulated-Raman transitions, we chose a qubit with an excited electronic state ${\ket{\uparrow}=\ket{^{2}S_{1/2},F=2,M_{F}=2}}$ and a ground electronic state ${\ket{\downarrow}=\ket{^{2}S_{1/2},F=3,M_{F}=3}}$. We use the IC system to perform resolved sideband cooling \cite{Didi2003}, cooling one motional mode of a Doppler cooled trapped ion close to the quantum ground state of the trap. The frequency difference between the two Raman beams is set to the red-sideband (RSB) frequency, the $\ket{\uparrow}\leftrightarrow \ket{\downarrow}$ transition frequency \emph{minus} the frequency of the motional mode being cooled. A RSB pulse promotes transitions from $\ket{\downarrow,n}\leftrightarrow \ket{\uparrow,n-1}$, where \emph{n} is the quantum number describing the motional state of the trapped ion \cite{Didi2003}. Repumping then brings the qubit to the ground electronic state $\ket{\downarrow,n-1}$ with one less motional quantum. This sequence is repeated until the ion reaches the $\ket{\downarrow,n=0}$ state. If the ion is close to the ground state, excitation of the ion at the red-sideband frequency is strongly suppressed, whereas excitation of the $\ket{\downarrow,n}\leftrightarrow\ket{\uparrow,n+1}$ blue-sideband (BSB) transition is not. Figure \ref{fig:Side_band_cooling} shows frequency scans of the first red and blue sidebands for a radial motional mode of the trap, after ground-state cooling using the IC system. Assuming that the motional states have a thermal distribution after cooling, the average motional state occupation number $\braket{n}$ of a given motional mode can be calculated from the ratio of the heights of the RSB and BSB peaks for that mode \cite{Didi2003}. From the data in Figure \ref{fig:Side_band_cooling} we calculate $\braket{n}=0.008\,^{+\,0.013}_{-\,0.008}\,$, consistent with this radial mode being in the motional ground state with high probability. Rabi flopping on the $\ket{\uparrow,n}\leftrightarrow \ket{\downarrow,n}$ carrier transition of the qubit, after ground state cooling, is shown in Figure \ref{fig:Flopping}. For this experiment, we obtain a Rabi frequency of 191.1$\,\pm\, 0.1$kHz and a $1/e$ decay time of 77$\,\pm \, 3\, $µs. The decoherence is due to a combination of off-resonant photon scattering, magnetic field noise and trap surface related heating sources.

\section{Conclusion}
We have developed an optically pumped VECSEL system for generating the wavelengths required for producing, cooling and manipulating magnesium ions. In the future, by use of an 1118$\,\rightarrow\,$559$\,\mathrm{nm}$ frequency doubling stage with a higher optical damage threshold, our system should be capable of generating sufficient UV power for achieving fault-tolerant, two-qubit quantum information gates with stimulated-Raman transitions \cite{Ozeri2007}. This could be accomplished by intracavity doubling, as in our VC laser setup, or through a more efficient external doubling step \cite{Friedenauer2006}. The power scalability of the VECSEL architecture makes it a promising candidate for systems that require individual laser addressing of multiple quantum systems. Due to their large wavelength tuning range, narrow linewidth, and low intensity noise, VECSELs systems have the potential to significantly reduce the complexity and cost of laser systems used for experiments in atomic and molecular physics.  

\section*{Funding Information}

This work was supported by the Academy of Finland project Qubit (decision number 278388) and TEKES project ReLase (40016/14) and by Office of the Director of National Intelligence (ODNI) Intelligence Advanced Research Projects Activity (IARPA), ONR, and the NIST Quantum Information Program. This paper is a contribution by NIST and not subject to US copyright.
\section*{Acknowledgments}
We thank Dr. Jeff Sherman for assistance with the three-cornered hat data collection and analysis. We also thank Dr. David Hume and Dr. Yong Wan for comments on the manuscript.
\bibliography{sample3}


\end{document}